\documentclass{aa}
\usepackage{txfonts}
\usepackage{epsfig}
\usepackage{subfigure}
\usepackage{multirow}

\usepackage{times}
\usepackage{natbib}
\usepackage{longtable,lscape}
\bibpunct{(}{)}{;}{a}{}{,}

\begin{document}

\title{The broad-band X-ray spectrum of the Seyfert 1 galaxy, MCG+8-11-11}
\author{Stefano Bianchi\inst{1}, Ilaria De Angelis\inst{1}, Giorgio
Matt\inst{1}, Valentina La Parola\inst{2}, Alessandra De Rosa\inst{3}, Paola
Grandi\inst{4}, Elena Jim\'enez Bail\'on\inst{5},
Enrico Piconcelli\inst{6}}

\offprints{Stefano Bianchi\\ \email{bianchi@fis.uniroma3.it}}

\institute{Dipartimento di Fisica, Universit\`a degli Studi Roma Tre, via della
Vasca Navale 84, 00146 Roma, Italy
\and INAF, Istituto di Astrofisica Spaziale e Fisica Cosmica, Via U. La Malfa
153, I-90146 Palermo, Italy
\and INAF/IASF-Roma, Via del Fosso del Cavaliere, I-00133 Roma, Italy
\and INAF-IASF Bologna, Via Gobetti 101, I-40129, Bologna, Italy
\and Instituto de Astronom\'ia, Universidad Nacional Aut\'onoma de M\'exico,
Apartado Postal 70-264, 04510 Mexico DF, Mexico
\and Osservatorio Astronomico di Roma (INAF), Via Frascati 33, I-00040 Monte
Porzio Catone, Italy}

\date{Received / Accepted}

\authorrunning{S. Bianchi et al.}

\abstract
{}
{Mounting evidence is showing that the main ingredients of the Unification
Models of Active Galactic Nuclei may behave differently from what expected, and
they could be intimately related to fundamental physical parameters. The
availability of high signal-to-noise broad-band X-ray spectra gives us the
opportunity to study in detail all the contributions from the materials invoked
in these models, and infer their general properties, including if their
presence/absence is related to other quantities.}
{We present a long (100 ks) \textit{Suzaku} observation of one of the X-ray
brightest AGN, MCG+8-11-11. These data were complemented with the 54-month
\textit{Swift} BAT spectrum, allowing us to perform a broad-band fit in the
0.6-150 keV range.}
{The fits performed in the 0.6-10 keV band give consistent results with respect
to a previous XMM-\textit{Newton} observation, i.e. the lack of a soft excess,
warm absorption along the line of sight, a large Compton reflection component
($R\simeq1$) and the absence of a relativistic component of the neutral iron
K$\alpha$ emission line. However, when the PIN and \textit{Swift} BAT data are
included, the reflection amount drops significantly ($R\simeq0.2-0.3$), and a
relativistic iron line is required, the latter confirmed by a phenomenological
analysis in a restricted energy band (3-10 keV). When a self-consistent model is
applied to the whole broadband data, the observed reflection component appears to be all
associated to the relativistic component of the iron K$\alpha$ line.}
{The resulting scenario, though strongly model-dependent, requires that all the reprocessing spectral components from
Compton-thick material must be associated to the accretion disc, and no evidence
for the classical pc-scale torus is found. The narrow core of the neutral iron
K$\alpha$ line is therefore produced in a Compton-thin material, like the BLR,
similarly to what found in another Seyfert galaxy, NGC~7213, but with the
notable difference that MCG+8-11-11 presents spectral signatures from an
accretion disc. The very low accretion rate of NGC~7213 could explain the lack
of relativistic signatures in its spectrum, but the absence of the torus in both
sources is more difficult to explain, since their luminosities are comparable,
and their accretion rates are completely different.}
{}

\keywords{Galaxies: active - Galaxies: Seyfert - quasars: general - X-rays:
general}

\maketitle

\section{Introduction}

The X-ray spectrum of unobscured AGN is characterised by ubiquitous and,
generally, not variable features, like the neutral iron K$\alpha$ narrow core
and the Compton reflection component \citep[e.g.][]{per02,bianchi07}, whose
origin appears quite well established within the framework of standard
Unification Models as the pc-scale `torus' \citep{antonucci93}. On the other
hand, there are other spectral components, like the warm absorber, the soft
excess, and the relativistic component of the iron K$\alpha$ line, which are
present only in a fraction of the sources, and may vary in different
observations of the same object \citep[e.g.][]{pico05,nan07}. Their origin is
less clear, and, like in the case of the relativistic iron line, they may pose
some crucial problems to the standard views.

X-ray emission from the accretion disc, as a result of reprocessing of the X-ray
photons produced in the corona, is a fundamental ingredient of any AGN model,
and should be characterised by a strong iron K$\alpha$ line, whose profile can
be clearly distinguished from the narrow core, because of well-known
relativistic effects occurring near the black hole \citep[see e.g.][]{fab00}.
However, this relativistic profile, when large samples of AGN are systematically
analysed, is not observed in all the sources \citep[see e.g.][]{nan07,long08a}.
Although at least a part of this problem must be related to the signal-to-noise
of the available X-ray spectra (indeed the fraction of sources with relativistic
signatures appears to be larger when only sufficiently exposed X-ray spectra are
analysed: see, e.g., de la Calle et al., subm.), the relativistic component of
the iron K$\alpha$ line still appears not to be ubiquitous, and, even when
present, its equivalent width (EW) is often lower than expected. 

These results impose the revision of standard models of accretion discs.
Truncation of the disc, highly ionization state, and complex illumination may
all be invoked to explain the observed deviations from the current picture, but
the fundamental issue is to understand the ultimate drivers for this behaviour,
like the accretion rate, the luminosity, and the black hole (BH) mass. In this
respect, mounting evidence is showing that also other ingredients of the
Unification Models may behave differently from what expected, and their own
presence could be intimately related to the same physical parameters. For
example, the broad line region (BLR) was found to be likely absent in some
sources \citep[e.g.][]{bianchi08b,panessa09}, acting as a Compton-thick X-ray
absorber in other \citep[e.g.][]{elvis04, ris05, puc07, bianchi09c}, or even
producing most of the observed neutral iron K$\alpha$ line, when no other
evidence for the presence of the torus is found in the data \citep{bianchi08}.

MCG+8-11-11 (z=0.0205) is one of the brightest AGN in the X-ray band, detected
by \textit{INTEGRAL} with a $F_\mathrm{20-100\,keV}\simeq1.2\times10^{-10}$ cgs
\citep{beck09}. \textit{ASCA} \citep{grandi98} and BeppoSAX \citep{per00,per02}
found that the spectrum is well fitted by a pretty standard model composed of a
power law, a warm absorber, a Compton reflection component, and an iron
K$\alpha$ line. The higher quality of the XMM-\textit{Newton} spectrum
\citep{matt06} revealed some interesting features: the lack of a soft excess, a
large reflection component, together with a narrow iron line with a low EW.
Moreover, the presence of a large, relativistically broadened line arising from
the accretion disk could be excluded in these data.

In this paper, we present a long (100 ks) \textit{Suzaku} observation of
MCG+8-11-11. Together with the 54-month \textit{Swift} BAT spectrum, we explore
different energy bands, in order to understand the origin of the X-ray spectral
components of this bright Seyfert 1. The availability of a high signal-to-noise
spectrum in such a broad band (0.6-150 keV) gives us the opportunity to study in
detail all the contributions from the materials invoked in Unification Models,
and infer their presence/absence and general properties.

\section{\label{datared}Data reduction}

MCG+8-11-11 was observed by \textit{Suzaku} on 2007, September 17, for 100 ks (\textsc{obsid} 702112010).
X-ray Imaging Spectrometer (XIS) and Hard X-ray Detector (HXD) event files were
reprocessed with the latest calibration files available (2010-03-23 release),
using \textsc{ftools} 6.9 and \textsc{Suzaku} software Version 16, adopting
standard filtering procedures. Source and background spectra for all the three
XIS detectors were extracted from circular regions with radius of 250 pixels
($\simeq260$ arcsec), avoiding the calibration sources. Response matrices and
ancillary response files were generated using \textsc{xisrmfgen} and
\textsc{xissimarfgen}. We downloaded the “tuned” non-X-ray background (NXB) for
our HXD/PIN data provided by the HXD team and extracted source and background
spectra using the same good time intervals. The PIN spectrum was then corrected
for dead time, and the exposure time of the background spectrum was increased by
a factor of 10, as required. Finally, the contribution from the cosmic X-ray
background (CXB) was subtracted from the source spectrum, simulating it as
suggested by the HXD team.

\begin{figure}
\epsfig{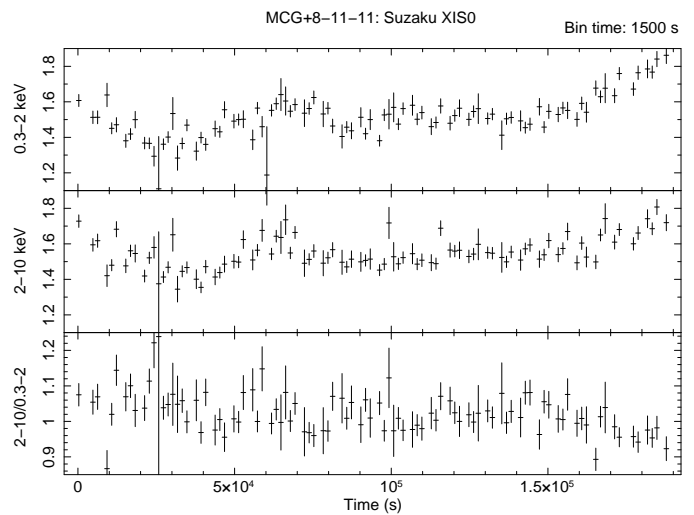}
\caption{\label{lcurve}Soft (0.3-2 keV) and hard (2-10 keV) light-curves of the
\textit{Suzaku} 100 ks observation of MCG+8-11-11 (only the XIS0 is plotted for
clarity). The bottom panel shows the ratio between the two curves.}
\end{figure}

We extracted the light-curves of the three XIS in the soft and the hard energy
band (see Fig.~\ref{lcurve} for the XIS0). The source presents some variability
in both bands, but the hardness ratio is substantially constant within 10\%. On
the other hand, the HXD pin light-curve does not show any significant evidence
of variability. Therefore, we decided to analyse the total spectra, without any
further timing analysis.

The front-illuminated XIS0 and XIS3 were fitted between 0.6 and 10 keV, while
the back-illuminated XIS1 was used up to 8 keV, because the background above
that energy becomes significant with respect to the source spectrum for that
instrument. The band 1.6-2.1 keV was excluded from all the fits, because of
known inter-calibration issues at those energies due to the Si K edge. The
normalizations of XIS1 and XIS3 with respect to XIS0 were left free in the fits,
and always resulted to be in agreement within 2\%. A constant factor of 1.18 was
instead used between the PIN and the XIS0, as recommended for observations taken
in the HXD nominal position. In the following, all the PIN fluxes will be given
with respect to the XIS0 flux scale, which is 1.18 times lower than the HXD absolute flux scale.

In the following, errors correspond to the 90\% confidence level for one
interesting parameter ($\Delta\chi^2=2.71$), where not otherwise stated. The
adopted cosmological parameters are H$_{0}=70$ km s$^{-1}$ Mpc$^{-1}$,
$\Omega_\Lambda=0.73$, and $\Omega_m=0.27$ \citep[i.e., the default ones in
xspec 12.6.0:][]{xspec}. We use the \citet{ag89} abundances and the
photoelectric absorption cross-sections by \citet{bcmc92}.

\section{Data analysis}

\subsection{\label{iron}The iron line complex}

\begin{figure*}
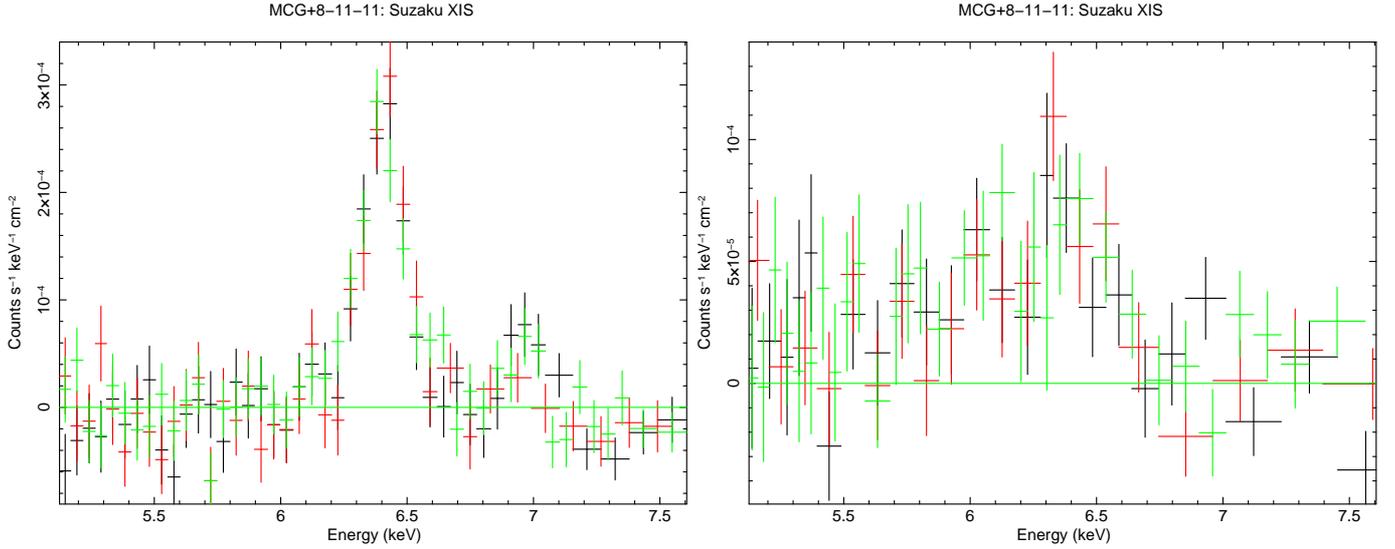

\epsfig{file=felines.ps,angle=-90,width=\columnwidth}
\epsfig{file=felines_ky.ps,angle=-90,width=\columnwidth}
\caption{\label{felines}\textit{Left}: residuals in the (rest-frame) iron line band, after a
3-10 keV fit on the three \textit{Suzaku} XIS spectra, with a simple power law. \textit{Right}: the same as above, but for model
III, after the relativistic component of the iron line is removed (see Table~\ref{felines_fits}). Data have been re-binned for clarity.}
\end{figure*}

As a first step we fitted the 3-10 keV spectra with a simple power law. The
photon index is quite flat ($1.62\pm0.02$), and clear residuals are left between
6 and 7 keV (see Fig.~\ref{felines}). Three Gaussian lines are required by the
data, at $6.401\pm0.008$, $6.66\pm0.06$, and $6.96\pm0.02$  keV, consistent with
emission from neutral, He-like and H-like Fe K$\alpha$. The presence of an
absorption edge from neutral iron \citep[its energy fixed to 7.11 keV:][]{bea67}
is also  marginally detected, with an optical depth $\tau=0.04\pm0.02$ (model
I). On the other hand, a Fe K$\beta$ emission line is not required by the data,
with an upper limit of $\simeq7\%$ with respect to the flux of the Fe K$\alpha$.
Analogously, the K$\alpha$ emission from nickel is not detected. While the two
ionised iron lines are consistent with being unresolved ($\sigma<80$ eV), the
neutral Fe K$\alpha$ is resolved, with $\sigma=52^{+17}_{-19}$ eV. This width is
larger than the calibration uncertainties of the XIS redistribution
matrix\footnote{We extracted the calibration spectra for this observation, and
fitted them separately for each XIS with a model consisting of three emission
lines: the Mn K$\alpha$ doublet \citep[the separation of the lines fixed to be
0.0111 keV:][]{bea67} and the K$\beta$. The width of the three lines were kept
linked. We recovered upper limits in all the spectra: 23 eV (XIS0), 25 eV
(XIS1), 14 eV (XIS3)}, and corresponds to a full width at half maximum
FWHM=$6500\pm1900$ km s$^{-1}$, if due to Doppler broadening. 

The neutral iron K$\alpha$ width can be explained by the contribution of a
Compton Shoulder (CS), as suggested by the presence of some residuals redwards
the line. We therefore added a further Gaussian line, with centroid at 6.3 keV
and $\sigma=40$ eV \citep[both fixed: see][]{matt02}. The improvement of the fit
is marginal (model II: $\Delta\chi^2=4$), but the Fe K$\alpha$ is now unresolved
($\sigma<45$ eV). However, its centroid is shifted to $6.421^{+0.011}_{-0.012}$
keV, and the flux ratio between the CS and the core is $32^{+10}_{-14}$\%,
larger than the theoretical expectations in Compton-thick matter \citep{matt02}.

Alternatively, the width of the iron line and the redwards residuals may be
signatures of a mildly relativistic line. We tested this scenario adding to the
unresolved core a \textsc{kyrline} model component \citep{dky04}, adopting a
fixed centroid energy at 6.4 keV, an outer radius of 400 r$_g$, and a power law
index $\beta=3$ for the radial dependence of the emissivity.
This fit (model III) is statistically preferred to the others ($\Delta\chi^2=15$ and 11 with respect to model I and II, and a null hypothesis probability of 0.43, significantly better than 0.22 and 0.26, respectively). 
The inner radius ($<20$r$_g$) and the inclination angle ($\theta=30\pm3\deg$) can be constrained in the
fit, while the BH spin is completely unconstrained, and therefore fixed to zero.
The Fe\textsc{xxv} line is not required by the data anymore: if the centroid
energy is allowed to vary between or fixed to the theoretical values for each component of this multiplet
\citep[see e.g.][]{bianchi05}, only upper limits are recovered (of the order of $1\times10^{-5}$ ph cm$^{-2}$ s$^{-1}$). The iron K edge
is also an upper limit in this model ($\tau<0.03$). The final fit is very good
($\chi^2=301/298$ d.o.f.), and will be applied in the following broad-band fits,
where the presence of the relativistic component will be further investigated
once a more sophisticated model for the continuum is adopted.
Table~\ref{felines_fits} summarizes the best fit parameters for the models
described in this section.

\begin{table}
\caption{\label{felines_fits}Best fit parameters for the emission features in
the \textit{Suzaku} XIS spectra (3-10 keV). The three models are described in
the text.}
\begin{tabular}{ccccc}
 & & \textbf{I} & \textbf{II} & \textbf{III} \\
&&&&\\
\multirow{4}{*}{Fe K$\alpha$}& $E_0$ & $6.401\pm0.008$ &
$6.421^{+0.011}_{-0.012}$ & $6.396^{+0.016}_{-0.004}$\\
& $\sigma$ & $52^{+17}_{-19}$ & $<45$ & $0^*$\\
& $F$ & $6.6\pm0.6$ & $5.0\pm0.5$ & $4.9\pm0.5$\\
& EW & $89\pm8$ & $67\pm7$ & $67\pm7$\\
&&&&\\
Fe K$\alpha$ CS& $F$ & -- & $1.6^{+0.5}_{-0.7}$ & --\\
&&&&\\
\multirow{3}{*}{{Fe\,\textsc{xxv}} K$\alpha$}& $E_0$ & $6.65\pm0.07$ &
$6.63^{+0.05}_{-0.07}$ & --\\
& $F$ & $0.6\pm0.4$ & $0.8^{+0.4}_{-0.3}$ & --\\
& EW & $8\pm5$ & $11^{+6}_{-4}$ & --\\
&&&&\\
\multirow{3}{*}{{Fe\,\textsc{xxvi}} K$\alpha$}& $E_0$ & $6.97\pm0.03$ &
$6.970^{+0.018}_{-0.032}$ & $6.960^{+0.019}_{-0.022}$\\
& $F$ & $1.5\pm0.4$ & $1.5\pm0.3$ & $1.9\pm0.4$\\
& EW & $23\pm6$ & $24\pm5$ & $29\pm6$\\
&&&&\\
\multirow{4}{*}{Fe K$\alpha_{ky}$}& $\theta$ & -- & -- & $30\pm3\degr$\\
& $r_i$ & -- & -- & $<20$\\
& $F$ & -- & -- & $6.1^{+1.7}_{-1.5}$\\
& EW & -- & -- & $90\pm20$\\
&&&&\\
Fe K edge & $\tau$ & $0.04\pm0.02$ & $0.04\pm0.02$ & $<0.03$\\
&&&&\\
& $\chi^2$/dof & 316/297 & 312/297 & 301/298\\
&&&&\\
\end{tabular}
Energies ($E_0$) are in keV, EWs and $\sigma$ in eV, fluxes ($F$) in units of $10^{-5}$ ph
cm$^{-2}$ s$^{-1}$, and radii ($r_i$) in $r_g$.
\end{table}

\subsection{The 0.6-10 keV band}

The best fit adopted for the 3-10 keV band (model III in Table~\ref{felines_fits}) can be applied to the whole XIS band, once
Galactic absorption along the line of sight is taken into account. We recover a
neutral column density\footnote{This value of column
density was found adopting the \textsc{tbabs} model, and the abundances and
cross-sections minimizing the $\chi^2$ (see Sect.~\ref{datared} for the final
choice).} of $\simeq2.1-2.4\times10^{21}$ cm$^{-2}$, depending on
the adopted complete model (see below and Table~\ref{xisfits}).
This measure is in good agreement with the one estimated by \citet{dl90}, $2.09\times10^{21}$ cm$^{-2}$, but somewhat larger than the estimate made by \citet{kalb05}, $1.76\times10^{21}$ cm$^{-2}$. Fixing the Galactic column density to the latter value, a comparable $\chi^2$ is recovered by adding an additional neutral column density at the redshift of the source ($4.6^{+0.9}_{-1.4}\times10^{20}$ cm$^{-2}$), without affecting any other parameter of the fit.

However, there is clearly some curvature left in the residuals at low energies,
resulting in a $\chi^2=681/515$. Since additional neutral absorption at the redshift of the
source is not required by the data, we added a warm absorber component \citep[at
first we adopted \textsc{absori}, with the temperature of the material kept
fixed to $10^6$ K, which can be directly compared to the results reported
by][]{matt06}, improving our fit down to $\chi^2=641/513$, with best fit parameters for the warm absorber in good agreement with those found by \citet{matt06}. As already noted by these authors, a single-zone absorber cannot describe properly the data. Some residuals are
still left around 0.8 keV: they can be fitted by an edge, found at
$0.82^{+0.02}_{-0.03}$ keV, with $\tau=0.050\pm0.017$. The energy of the edge is
intermediate between K absorption from {O\,\textsc{vii}} (0.7393 keV) and
{O\,\textsc{viii}} (0.8714 keV), but inconsistent with both. We note here that
this feature is not even close to the absorption features observed in the
XMM-\textit{Newton} RGS spectra \citep{matt06}.

\begin{figure}
\epsfig{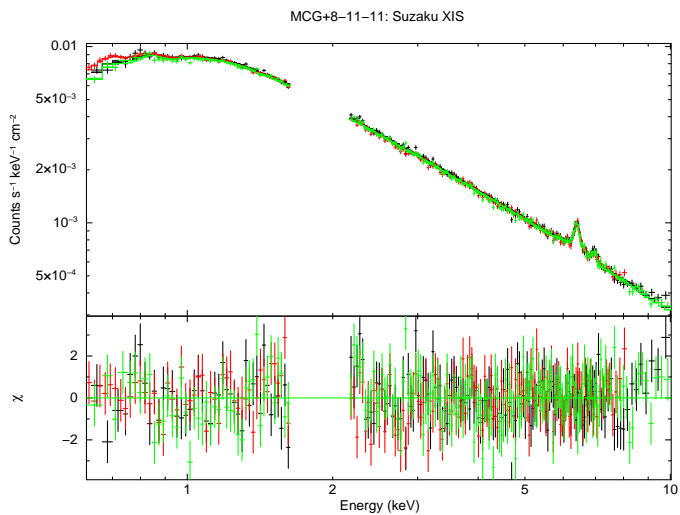}
\caption{\label{xisband}Best fit and $\Delta\chi^2$ for the three XIS spectra,
with the same model as in the 3-10 keV fit, but including Galactic absorption, a
warm absorber and an unidentified absorption edge (see text for details). Note
the residuals above 8 keV.}
\end{figure}

The resulting $\chi^2=616/511$ would be acceptable, since most of the residuals
appears to be dominated by bad subtraction of the background (as between 7 and 8
keV for the XIS0) and some inter-calibration issues between the instruments.
However, above 8 keV the data-points seem systematically higher than the model
(see Fig.~\ref{xisband}). We therefore added a Compton reflection component
(model \textsc{pexrav}, with inclination angle fixed to $30\degr$ and cutoff
energy to 150 keV, see next section), which indeed further improves the fit, for
a final value of $\chi^2=586/509$. The photon index is now steeper
($\Gamma=1.772\pm0.017$) than the one found in the previous section for the
limited band, and the amount of Compton reflection is R=$0.9\pm0.3$, with an
iron abundance of $A_{Fe}=0.6^{+0.3}_{-0.2}$. The need for a relativistic
component of the neutral iron line is much relaxed, given the large inner radius
($110^{+180}_{-80}$ r$_g$) and the low EW ($30\pm12$ eV). In order to understand
if the disappearance of the relativistic line and the other fitting parameters
strongly depend on our choice of a simple model for warm absorption, we tried a
more sophisticated warm absorber model with respect to \textsc{absori}, which
only takes into account absorption edges. 

\begin{table*}
\caption{\label{xisfits}Best-fits for the \textit{Suzaku} XIS spectra
(0.6-10 keV, first two columns) and the \textit{Suzaku}+\textit{Swift} BAT
spectra (0.6-150 keV, last two columns). The adopted model is the same in all the cases: \textsc{TBabs*zedge*absori*(kyrline+zgauss+zgauss+pexrav} in \textsc{xspec} jargon (see also model III in Table~\ref{felines_fits}). For each band, results are reported when the warm absorber is modelled with \textsc{absori} and \textsc{cloudy}. See text for details.}
\begin{center}
\begin{tabular}{ccccc}
& \multicolumn{2}{c}{\textbf{0.6-10 keV}} & \multicolumn{2}{c}{\textbf{0.6-150
keV}}\\
& \textsc{absori} & \textsc{cloudy} & \textsc{absori} &\textsc{cloudy}\\
$N_{Hg}$ ($10^{21}$ cm$^{-2}$) & $2.37\pm0.05$ & $2.22^{+0.09}_{-0.15}$ &
$2.32\pm0.04$ & $2.18^{+0.08}_{-0.11}$\\
$N_{Hw}$ ($10^{22}$ cm$^{-2}$) & $0.9^{+1.1}_{-0.6}$ & -- & $1.5^{+1.0}_{-0.6}$
& --\\
$\log N_{Hw}$ & -- & $20.7\pm0.2$ & -- & $20.5^{+0.3}_{-0.5}$\\
$\xi$ (erg s$^{-1}$ cm) & $600^{+900}_{-200}$ & -- & $600^{+370}_{-160}$ & --\\
$\log U$ & -- & $-1.2\pm0.6$ & -- & $-1.1^{+1.1}_{-0.8}$\\
$E_\mathrm{edge}$ & $0.827^{+0.015}_{-0.017}$ & $0.829\pm0.015$ & $0.82\pm0.02$
& $0.83\pm0.02$\\
$\tau_\mathrm{edge}$ & $0.072\pm0.018$ & $0.074^{+0.019}_{-0.020}$ &
$0.060\pm0.017$ & $0.060\pm0.018$\\
$\Gamma$ & $1.772\pm0.017$ & $1.80\pm0.02$ & $1.740\pm0.012$ & $1.743\pm0.012$\\
$R$ & $0.9\pm0.3$ & $1.4\pm0.3$ & $0.23\pm0.04$ & $0.27\pm0.05$\\
$E_{c}$ & $150^*$ & $150^*$ & $150^{+30}_{-20}$ & $150^{+30}_{-20}$\\
$A_{Fe}$ & $0.6^{+0.3}_{-0.2}$ & $0.73^{+0.21}_{-0.18}$ & $<0.33$ & $<0.15$\\
&&&&\\
$F_{0.5-2}$ & $2.15\pm0.02$ & $2.157\pm0.018$ & $2.151\pm0.017$ &
$2.14\pm0.03$\\
$F_{2-10}$ & $6.66\pm0.04$ & $6.66\pm0.03$ & $6.61\pm0.03$ & $6.60\pm0.03$\\
$F_{15-50}$ & -- & -- & $7.9\pm0.3$ & $8.1\pm0.2$\\
$L_{0.5-2}$ & $3.73\pm0.03$ & $3.59\pm0.03$ & $3.70\pm0.03$ & $3.64\pm0.05$\\
$L_{2-10}$ & $6.45\pm0.04$ & $6.41\pm0.03$ & $6.46\pm0.03$ & $6.38\pm0.03$\\
$L_{15-50}$ & -- & -- & $6.4\pm0.2$ & $6.48\pm0.16$\\
$\chi^2$/dof & 586/509 & 581/509 & 640/531& 658/531\\
&&&&\\
\end{tabular}
\end{center}
Fluxes are in $10^{-11}$ erg cm$^{-2}$ s$^{-1}$, unabsorbed luminosities in
$10^{43}$ erg s$^{-1}$. 15-50 keV fluxes and luminosities are relative to the XIS normalization (a factor of 1.18 lower than the HXD absolute flux).
\end{table*}

We prepared an ad-hoc table produced with the photoionization code
\textsc{cloudy} C08.00 \citep{cloudy}. The ingredients for this model are: a
multi-component continuum mimicking a typical AGN Spectral Energy Distribution
(SED), with $\alpha_{ox}=1.26$ and $\Gamma_{2-10\,\mathrm{keV}}=1.8$, as
extracted from the results of the simultaneous UV and X-ray data of the
XMM-\textit{Newton} observation of MCG+8-11-11 \citep{matt06}; constant electron
density\footnote{Models with higher electron densities, up to $n_e=10^9$ cm$^{-3}$ have also been produced, but all the following fits resulted insensitive to this parameter, as expected \citep[see e.g.][]{nic99}.} $n_e=10^4$ cm$^{-3}$; ionization parameter\footnote{We use the standard definition: $U=\frac {\int _{\nu _{R}}^{\infty }\frac{L_{\nu }}{h\nu} d\nu}{4\pi r^{2}cn_{e}}$ where $c$ is the speed of light, $r$ the distance of the gas from the illuminating source, $n_{e}$ the electron density and $\nu_{R}$ the frequency corresponding to 1 Rydberg \citep{of06}.} in the range $\log
U=-2.0:2.5$; intervening column density in the range $\log N_{NH}=20.0:22.0$. 
The resulting fit is marginally better than the one found with \textsc{absori}
($\Delta\chi^2=5$) and, notably, the unidentified absorption edge is still
required by the data, with parameters consistent with those found in the other
fit. However, the properties of the warm absorber gas are quite different, both
the column density and the ionization parameter being significantly lower in the
\textsc{cloudy} best fit, confirming that the signatures of warm absorption are weak in this source. Indeed, all the other best-fit parameters
are consistent with those found with \textsc{absori}, including the narrowish
and weak relativistic component of the Fe K$\alpha$ emission line. No improvement in the fit is found if a velocity shift is allowed for the warm absorbing gas, nor with the addition of a second warm absorber component.
The first two columns of Table~\ref{xisfits} summarize and compare the results for these two
fits.

\subsection{\label{broadbandfit}The broadband fit}

\begin{figure*}
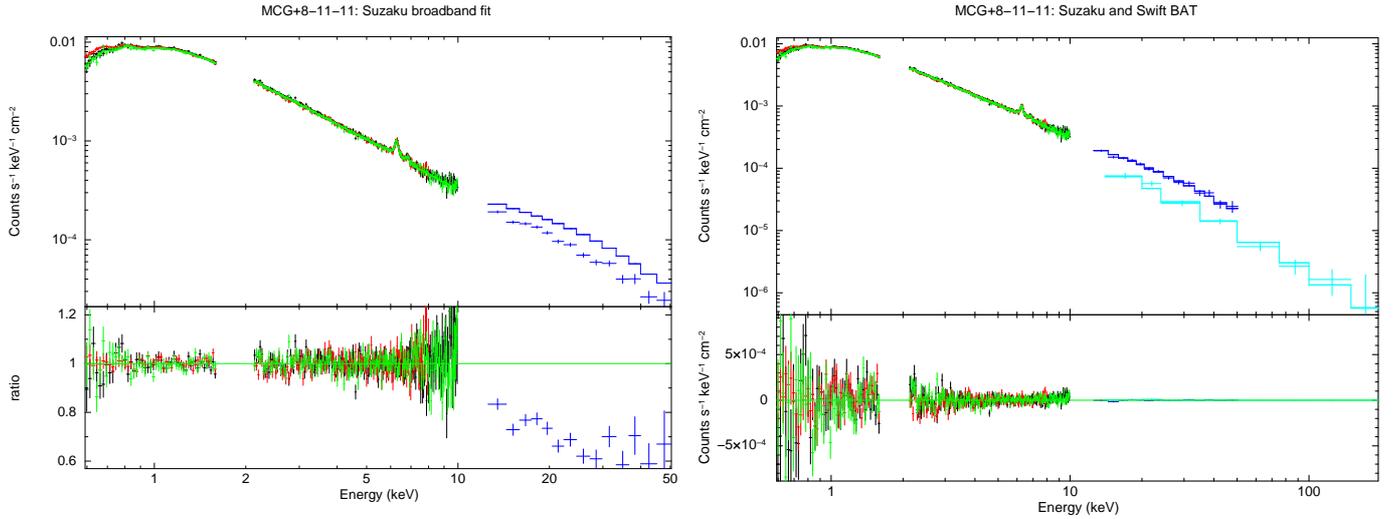

\epsfig{file=broadband_xisfit.ps,angle=-90,width=\columnwidth}
\epsfig{file=broadband_bat.ps,angle=-90,width=\columnwidth}
\caption{\label{broadband_xisfit}\textit{Left}: the best-fit model for the
0.6-10 keV band, when extrapolated to the band covered by the PIN spectrum.
\textit{Right}: the \textit{Suzaku}+\textit{Swift} BAT best-fit model in the
whole 0.6-150 keV band.}
\end{figure*}

When the best fit model found for the 0.6-10 keV XIS spectra is extrapolated to
higher energies, the PIN data points all fall well below the expectations (see
Fig.~\ref{broadband_xisfit}). Allowing the model to be adjusted for the
high-energy data, the best-fit is acceptable ($\chi^2=625/524$), the main
difference with the limited band fit being the very low value of the Compton
reflection component $R=0.13^{+0.08}_{-0.06}$. The relativistic component of the
iron K$\alpha$ line is now very significant, recovering the same properties
found in model III of the 3-10 keV fit (see Sect.~\ref{iron} and
Table~\ref{felines_fits}). The cutoff energy cannot be measured, with a $99\%$
lower limit of $\simeq130$ keV (for two interesting parameters, considering the
reflection amount as second interesting parameter). 

\begin{figure}
\epsfig{file=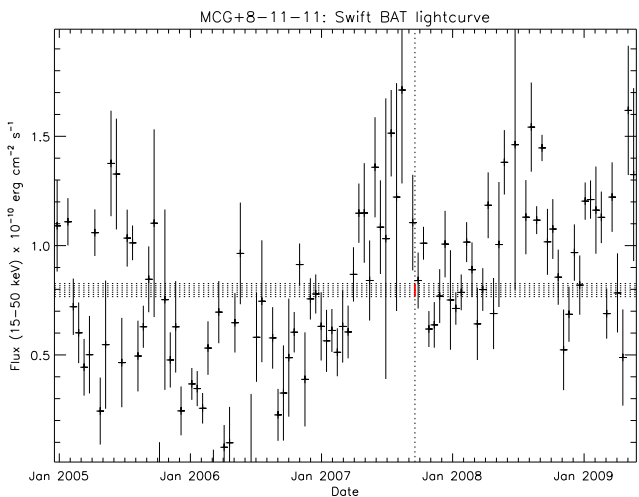,width=\columnwidth}
\caption{\label{batlightcurve}The 15-50 keV \textit{Swift} BAT 54-month lightcurve of
MCG+8-11-11 (a bin is 15 days). The countrate was converted to a 15-50 keV flux, assuming no spectral variation. The \textit{Suzaku} PIN flux is shown in comparison, with its
uncertainty, with the dotted lines and the red symbol. The agreement at the date
of the \textit{Suzaku} observation is good.}
\end{figure}

In order to constrain the cutoff energy, we then added the \textit{Swift} BAT
spectrum (in the range 15-150 keV), from the 54-Month Source Catalog (Cusumano
et al., submitted to A\&A). The normalization factor between the XIS0 and the BAT
spectrum was allowed to vary, in order to take into account source variability
during the 54 months of integration. Indeed, the best-fit value for this factor
($0.76\pm0.03$) clearly suggests that the \textit{Suzaku} observation caught the
source in a flux state significantly higher than the average of the BAT
campaign. This is confirmed by the BAT lightcurve, which show significant
variability in the 54 months, and a flux comparable to the one measured by the
PIN at the date of the \textit{Suzaku} observation (see
Fig.~\ref{batlightcurve}).
Once assumed that this flux variability does not
imply a spectral variability, we re-fitted the best fit model including the BAT spectrum, and found
$E_c=150^{+30}_{-20}$ keV. This value, still consistent with the lower limit
found with the \textit{Suzaku} data only, is also in agreement with what
measured by BeppoSAX \citep[$170^{+300}_{-80}$ keV:][]{per00} and only
marginally inconsistent with \textit{OSSE} \citep[$270^{+90}_{-70}$
keV:][]{grandi98}. 

\begin{figure*}
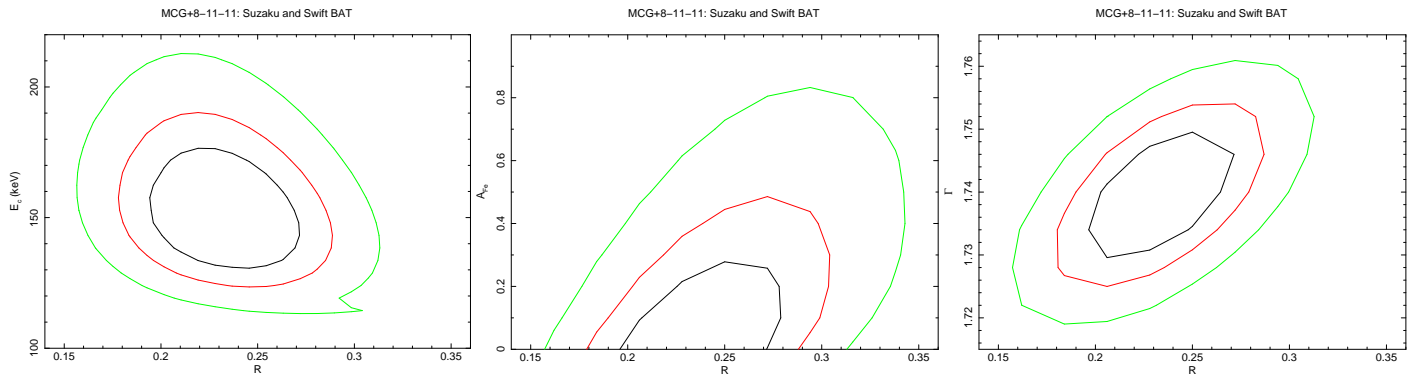

\epsfig{file=bat59months_R_foldE.ps,angle=-90,width=0.33\textwidth}
\epsfig{file=bat59months_R_Feabund.ps,angle=-90,width=0.33\textwidth}
\epsfig{file=bat59months_R_gamma.ps,angle=-90,width=0.33\textwidth}
\caption{\label{contours}\textit{Suzaku}+\textit{Swift} BAT best fit contour
plots (black, red, green lines correspond to 68, 90, and 99\% confidence level
for two interesting parameters). \textit{Left}: Compton reflection amount $R$ vs
cutoff energy $E_c$; \textit{Middle}: Compton reflection amount $R$ vs iron
abundance $A_{Fe}$; \textit{Right}: Compton reflection amount $R$ vs spectral
photon index $\Gamma$.}
\end{figure*}

Differently from the 0.6-10 keV fit discussed in the previous Section, the best
fit model for the broadband 0.6-150 keV spectra is achieved with the warm
absorber parametrised by \textsc{absori}: a significantly worse fit is found
when the \textsc{cloudy} table is used ($\Delta \chi^2=18$: see Table
\ref{xisfits}). In any case, all the other parameters appears to be independent of the
choice of the warm absorber. In particular, both fits recover a low reflection
component ($R\simeq0.2-0.3$) and an iron abundance significantly lower than the
Solar one. Their inter-correlation and dependencies upon other relevant
parameters are shown in Fig~\ref{contours}. Interestingly, the relativistic
component of the iron line is now very significant, and its best fit parameters
are similar to the ones found in our first phenomenological fits of the iron
line complex (see Section \ref{iron} and Table \ref{felines_fits}). 

The broadband best fit is not statistically as good as the one limited by the
XIS band-pass, being $\chi^2=640/531$. Most of the residuals are due to the XIS
data-points above 8 keV, which are on average higher than the model: as we have
seen in the previous section, they were likely the main drivers to require an
high Compton reflection in the 0.6-10 keV band, but they are apparently
inconsistent with the PIN flux. Alternatively, we tried to keep the iron
abundance fixed to the Solar value, leaving the inclination angle of the
reflecting material free in the fit. The resulting $\chi^2$ is comparable to the
best fit ($\chi^2=646/531$), with a very high inclination angle ($\cos i<0.30$).
The reflection fraction is now much larger, but loosely constrained
($R=2.9^{+0.6}_{-2.2}$).

\section{Discussion}

\subsection{Comparison with previous observations}

In the last 15 years, MCG+8-11-11 was observed in the X-rays by \textit{ASCA}
\citep[simultaneously with \textit{OSSE},][]{grandi98}, BeppoSAX \citep{per00},
and XMM-\textit{Newton} \citep{matt06}. Despite significant flux variability (a
factor $\simeq3$ between \textit{ASCA} and the present \textit{Suzaku}
observation), there is no clear evidence of variations of the spectral shape of
the source. The photon index of the primary continuum was also measured in the
range 1.7-1.8, a difference of 0.05 arising in this \textit{Suzaku} observation
whether the PIN data are considered or not in the fit. The warm absorber, if
modelled in the same way (i.e. \textsc{absori}) also gives consistent results
between XMM-\textit{Newton} and \textit{Suzaku}. The notable absence of a soft
excess is confirmed in all the observations, even if the large Galactic column
density may contribute to conceal it from the data.

The iron K$\alpha$ neutral line has a complex profile, and it is likely composed
of different components. Both \textit{ASCA} and BeppoSAX data suggested a
broadish profile ($\sigma\simeq0.2-0.3$ keV, but marginally resolved), with EWs
consistent with a constant line flux \citep{grandi98,per00}. The same applies
for the XMM-\textit{Newton} and the \textit{Suzaku} iron K$\alpha$ flux, once
taken into account all the components required by the data. However, while these
results suggest that the iron K$\alpha$ flux did not vary significantly, the
errors and the differences in modelling its profile make it difficult to extract
any other useful information from this apparent lack of response to the primary
flux variability. 

As for the Compton reflection component, all previous observations are
consistent with values around $R\simeq1-1.5$, but with a good (statistical)
constraint only in the XMM-\textit{Newton} data \citep{matt06}. The
\textit{Suzaku} best fit limited to the 0.6-10 keV band is consistent with these
values, but still in agreement, within the errors, with a scenario where the reflection component flux is
constant, and the value of $R$ anti-correlates with the flux of the primary
continuum, as expected if the reprocessing originates in a distant material,
like the torus. However, when the PIN data are taken into account, the Compton
reflection drops to a significantly lower value, around 0.2-0.3 (see Table
\ref{xisfits}). This is also supported by the preference for very low iron
abundance and/or large inclination angles with respect to the line-of-sight,
likely due to the fact that the iron edge is not particularly deep, as further
confirmed by the phenomenological fit we performed in the 3-10 keV band (see
Section \ref{iron}), where the edge is significant only when the relativistic
component of the iron K$\alpha$ line is not included in the model. It is
interesting to note that an iron abundance lower than Solar is also preferred in
the 0.6-10 keV fit, as found by \citet{matt06} in the XMM-\textit{Newton} data.
In these cases, though, the relativistic component of the iron K$\alpha$ line is
completely negligible.

\subsection{A global scenario}

In order to test the data with a self-consistent global scenario, we replaced in the model described in Sect.~\ref{broadbandfit} (see also Table~\ref{xisfits}) the \textsc{pexrav} and the narrow core of the iron K$\alpha$ line with the \textsc{pexmon} model \citep{nan07}, which includes at the same
time the incident primary power law, and all the reprocessing components arising
from a Compton-thick material: the emission features (Fe K$\alpha$, K$\beta$ and
Compton shoulder, and the Ni K$\alpha$ emission line) and the Compton reflection
hump, with all the relevant inter-dependencies (inclination, index and
abundance). The resulting fit is worse than the best fit, but still acceptable
($\chi^2=650/532$). The iron abundance is now much higher
($A_{Fe}=1.1^{+0.3}_{-0.2}$), and the amount of reflection is somewhat larger
($R=0.34^{+0.05}_{-0.04}$). This is due to the need to model the narrow
component of the iron neutral K$\alpha$ line self-consistenly with the Compton
reflection hump. On the other hand, all the other parameters are consistent with
those found in our best fit presented in the previous section, within the
errors, including the warm absorber (\textsc{absori}) and the absorption edge. 

We then tested if a further component to the neutral iron line narrow core is
needed by the data. A possibility is a contribution from the BLR. In the case of
MCG+8-11-11, the FWHM of the H$\alpha$ is 2920 km s$^{-1}$, and that of the
H$\beta$ 3630 km s$^{-1}$ \citep{os82}. We therefore added a neutral Fe
K$\alpha$ line with $\sigma$ fixed at 30 eV (consistent with a production in the
same gas where H$\alpha$ and H$\beta$ originate), and fixed the iron abundance
at the Solar value. The fit only slightly improves ($\chi^2=647/532$), due to
the marginal significance of the iron K$\alpha$ line supposedly produced in the
BLR (EW=$14\pm12$ eV). This test seems to suggest that the observed iron
K$\alpha$ line can be satisfyingly accompanied by the measured low reflection
component, in a self-consistent way. But this is due to the fact that only the
narrow core needs to be modelled, with EW=60-70 eV (see Table~\ref{felines}).

Indeed, in all these fits, a significant contribution from a relativistic
component to the iron line is still required. This component should be
accompanied by its own Compton reflection, produced in the accretion disc. To
test this scenario, we therefore added another \textsc{pexmon} component (only
the reprocessing components), convolved with a relativistic kernel
(\textsc{kdblur}) to model the relativistic effects occurring in the inner
accretion disc, close to the BH. The final best fit further improves, and it is
almost statistically equivalent to the best fit ($\chi^2=644/532$). All the
reflection component is due to the disc, with $R=0.31^{+0.03}_{-0.14}$, while
the contribution from the torus is negligible ($R<0.1$). Therefore, all the
narrow component of the iron K$\alpha$ line is now modelled by emission from the
BLR ($\sigma=30$ eV as above), with an EW$\simeq70$ eV. The inner radius of the
accretion disc is just beyond the innermost stable orbit for a non-spinning BH
($13^{+31}_{-4}$ r$_g$).

The resulting scenario would be very interesting. If the reflection component
must be all ascribed to the accretion disc, consistently with the observed
relativistic component of the iron K$\alpha$ line, it is unavoidable that the
narrow core of the emission line is produced in a Compton-thin material, like
the BLR, whose contribution to the reflection amount would be negligible. This,
in turns, implies that there would be no evidence for the classical
Compton-thick torus in this source. Given the high-flux state of the source during the \textit{Suzaku} observation (see Sect.~\ref{broadbandfit} and Fig.~\ref{batlightcurve}), it is possible that the reprocessing components from the torus are diluted by the stronger primary continuum. However, this would justify the low (or absent) Compton reflection from the torus, but most of the narrow iron K$\alpha$ line should still be produced in a Compton-thin material, because the component arising from the torus would be diluted, too.

This situation is similar to the one
observed in another Seyfert galaxy, NGC~7213, the only Seyfert 1 galaxy with a
negligible amount of Compton reflection \citep{bianchi03b,bianchi08,lob10}. The
observed iron neutral line in this source ($EW\simeq100$ eV) was explained by
\citet{bianchi08} as being produced mainly in the BLR, as confirmed by the
consistency between the FWHM of the broad component of the H$\alpha$ line and
the neutral iron K$\alpha$ line, in quasi-simultaneous optical/\textit{Chandra}
observations. The same scenario would apply for MCG+8-11-11, but with the
notable difference that this object presents spectral signatures from an
accretion disc.

We found two BH mass estimates for MCG+8-11-11 in the literature, both derived by means of the BLR radius vs 5100
$\AA$ relationship, and the H$\beta$, but with different data: $1.5\times10^7$ M$_\odot$
\citep{bz03} and $1.2\times10^8$ M$_\odot$ \citep{win10}. Adopting a bolometric correction of 30 to the observed 2-10 keV
luminosity \citep[as appropriate in this luminosity range, according
to][]{mar04}, we recover $L_{bol}/L_{edd}\simeq0.1-1$. Although a better BH mass
estimate is needed to assess the reliability of this accretion rate,
it is clear the difference with respect to NGC~7213, characterised by a low
accretion rate \citep[$\simeq3\times10^{-3}$:][]{bianchi09}. This could explain
the lack of relativistic signatures in the latter, where the accretion disc may
be truncated and its inner part replaced by some sort of inefficient disc
\citep[see e.g.][]{lob10}. The absence of the torus in both sources is instead more
puzzling. The formation of the torus may be suppressed at low accretion rates
and/or luminosities, as predicted for the BLR \citep[e.g.][]{nic00,es06}.
However, as we have already seen, the accretion rates of these two sources have
nothing in common, and their luminosities are by no means low.

\section{Conclusions}

We presented a long \textit{Suzaku} observation of one of the X-ray brightest
AGN, MCG+8-11-11. The spectrum is characterised by a standard power-law photon
index of 1.7-1.8, and a warm absorber which cannot be satisfyingly modelled with
a single gas component. The notable absence of a soft excess, nearly ubiquitous in
unobscured Seyfert galaxies, may be partly due to the high Galactic column
density along the line-of sight.

The fits performed in the 0.6-10 keV band give consistent results with respect
to a previous XMM-\textit{Newton} observation, i.e. a large Compton reflection
component ($R\simeq1$) and the absence of a relativistic component of the
neutral iron K$\alpha$ emission line. However, when the PIN data are included,
the reflection amount drops significantly ($R\simeq0.2-0.3$), and a relativistic
iron line is required, the latter confirmed by a phenomenological analysis in a
restricted energy band (3-10 keV). The addition of the \textit{Swift} BAT
54-month spectrum up to 150 keV allowed us to measure an high-energy cutoff at
150 keV, in agreement with previous results.

When a self-consistent model is applied to the whole data, the observed
reflection component appears to be all associated to the relativistic component
of the iron K$\alpha$ line. Therefore, all reprocessing from a Compton-thick
material must be associated to the accretion disc, and no evidence for the
classical pc-scale torus is found. It is unavoidable that the narrow core of the
neutral iron K$\alpha$ line is produced in a Compton-thin material, like the
BLR, whose contribution to the reflection amount would be negligible. Although more model-dependent, this
situation is similar to the one observed in another Seyfert galaxy, NGC~7213,
the only Seyfert 1 galaxy with a negligible amount of Compton reflection, but
with the notable difference that MCG+8-11-11 presents spectral signatures from
an accretion disc. The very low accretion rate of NGC~7213 could explain the
lack of relativistic signatures in its spectrum, but the absence of the torus in
both sources is more difficult to explain, since their luminosities are
comparable, and their accretion rates are completely different.

\acknowledgement
This research has made use of data obtained from the \textit{Suzaku} satellite,
a collaborative mission between the space agencies of Japan (JAXA) and the U.S.A. (NASA).
We thank the anonymous referee for useful suggestions which improved the clarity of the paper.
SB, EP and GM acknowledge financial support from ASI (grant I/088/06/0). We
would like to thank K. Nandra, G. Miniutti, K. Arnaud and C. Gordon for the
\textsc{xspec} model \textsc{pexmon}, and K. Hamaguchi for his support on
\textit{Suzaku} data reduction and analysis.

\bibliographystyle{aa}
\bibliography{sbs}

\end{document}